\documentclass[12pt]{article}
\usepackage{exscale,epsfig,a4}
\begin{document}
\title{\bf Quantum state conversion\\
           by cross-Kerr interaction}
\author{J Clausen, L Kn\"oll and D-G Welsch
\\[3ex]
         Friedrich-Schiller-Universit\"at Jena,\\
         Theoretisch-Physikalisches Institut,\\
         Max-Wien-Platz 1, D-07743 Jena, Germany,\\
         Fax: ++49 (0)3641 9 47102,\\
         Email: J.Clausen@tpi.uni-jena.de
\\[1ex]
       } 
\date{}
\maketitle
\begin{abstract}
A generalized Mach--Zehnder-type interferometer equipped with
cross-Kerr elements is proposed to convert $N$-photon truncated
single-mode quantum states into ($N$+1)-mode single-photon states,
which are suitable for further state manipulation
by means of beam splitter arrays and ON/OFF-detections,
and vice versa.
Applications to the realization of unitary and non-unitary transformations,
quantum state reconstruction, and quantum telemanipulation are studied.
\end{abstract}


\newpage
%-----------------------------------------------------------------------------
\section{Introduction}
\label{sec1}
%-----------------------------------------------------------------------------
As a consequence of photon number conservation, the cross-Kerr interaction
offers an ideal playground for quantum state engineering, and a number of
applications have been studied, such as quantum non-demolition measurement,
quantum state preparation and detection, quantum teleportation, and the
implementation of quantum logic gates. The following examples reflect this
variety of fields touched upon.
A theoretical study of quantum non-demolition measurement of the photon number
of two optical modes based on cross-Kerr couplers in combination with a
Mach-Zehnder-interferometer is carried out in \cite{qndKim}.
In \cite{solitonSizmann}, an experimental review with emphasis on back-action
evading measurements on optical soliton pulses propagating in fibers is given.
The preparation of Schr\"odinger cat-like states by employing the cross-Kerr
interaction in conditional measurement is discussed in \cite{catVitali},
including a study of the effect of damping.
In \cite{prepAriano}, the use of a ring cavity equipped with a cross-Kerr
element and an ON/OFF-detector with arbitrary efficiency is proposed to
project a desired Fock state out of a coherent state, and in \cite{measAriano}
a chain of such ring cavities is used to measure the photon number statistics
of a signal.
In \cite{teleVitali}, a realization of a Bell state measurement based on the
cross-Kerr interaction is suggested and its application to teleporting the
polarization state of a photon investigated.
In \cite{switchAriano}, a fibre-optic non-linear Sagnac interferometer working
as an optical switch is analysed and its application as an optical regenerator
examined.
The implementation of a quantum phase gate operating on two polarization qubits
is considered in \cite{gateGlancy}.
Entanglement purification generating maximally entangled states from Gaussian
continuous entangled states by applying cross-Kerr interactions in conditional
measurement is studied in \cite{puriZoller}.

Within classical optics, the cross-Kerr interaction is described as a
third-order deviation from linearity of the polarization induced in a medium
by an electric field, so that strong fields are expected to be required for its
observation.
In contrast, setups discussed within quantum optics and quantum information
processing often operate with superpositions of low-excited Fock states.
An enhancement of the ``classical'' nonlinearity is therefore required for its
applicability in the domain of weak fields.
Quantum effects such as electromagnetically induced transparency could offer a
way to realize this enhancement \cite{giantSchmidt1}-\cite{giantWelsch}.
An alternative are proposals entirely based on the
nonlinearity hidden in the quantum measurement process \cite{linearMilburn}.

In the present article we assume an interaction of the form
$\hat{K}=\exp(\mathrm{i}\kappa\hat{n}_1\hat{n}_0)$ with 
\mbox{$\kappa$ $\!\in$ $\![0,2\pi]$} and show that it can be used,
in combination with beam splitter arrays and ON/OFF-detectors, to convert
$N$-photon truncated single-mode quantum states into ($N$+1)-mode single-photon
states and vice versa.
Such a converter offers novel possibilities of arbitrary single-mode quantum
state engineering. As potential applications, we consider the realization of
unitary and non-unitary operators, overlap measurements with orthogonal and
non-orthogonal sets of states \cite{nonorthHillery}, and quantum
telemanipulation. 
In general, it should be noted that finite dimensional quantum systems play an
important role in the study of basic quantum state engineering and detection
techniques \cite{nonorthHillery}-\cite{discretBjoerk}.

An ON/OFF-detector is a photodetector able to distinguish between presence and
absence of photons and may be realized by an avalanche-triggering photodiode.
Since the total photon number of single-photon states is one, the detection of
presence of photons can be done with any (non-zero) quantum detection
efficiency. Placing emphasis on the main principle, we will however
assume unit detection efficiency throughout the work.

The article is organized as follows. After introducing quantum state conversion
in section~\ref{sec2}, a proposal of its practical implementation is made in
section~\ref{sec3}. Application to quantum-state engineering is considered in
section~\ref{sec4}. Sections \ref{sec5} and \ref{sec6} are devoted to
applications to quantum-state measurement and quantum telemanipulation
respectively. Finally, a summary and some concluding remarks are given in
section~\ref{sec7}.
%-----------------------------------------------------------------------------
\newpage\section{Quantum state conversion}
\label{sec2}
%-----------------------------------------------------------------------------
Let $\hat{\varrho}$ be an arbitrary quantum state in a source Hilbert space and
$\hat{Y}(\omega)$ the operator that, as a result of a measurement with outcome
$\omega$, converts $\hat{\varrho}$ into a state
\mbox{$\hat{\varrho}^\prime$ $\!=$ $\!\hat{\varrho}^\prime(\omega)$}
in an isomorphic target Hilbert space according to  
\begin{equation}
  \hat{\varrho}^\prime = \frac{1}{p(\omega)}
  \hat{Y}(\omega)\hat{\varrho}\hat{Y}^\dagger(\omega),
\label{yomega}
\end{equation}
where
\begin{equation}
  p(\omega)=
  \bigl\langle\hat{Y}^\dagger(\omega)\,\hat{Y}(\omega)\bigr\rangle
\label{pomega}
\end{equation}
is the corresponding success probability,
and the operators $\hat{Y}^\dagger(\omega)\,\hat{Y}(\omega)$
must resolve the identity in the source space,
\mbox{$\sum_\omega\hat{Y}^\dagger(\omega)\,\hat{Y}(\omega)$ $\!=$ $\!\hat{I}$}.

In what follows we restrict our attention to the source space $\mathcal{H}_a$
spanned by the $k$-photon single-mode states
\mbox{$|k\rangle$ $\!=$ $\!(k!)^{-1/2}\hat{a}^{\dagger k}|0\rangle_a$}, where
\mbox{$k$ $\!=$ $\!0,\ldots,N$}, and the target space $\mathcal{H}_b$ spanned
by the ($N$+1)-mode single-photon states defined by
\mbox{$|\varphi_k\rangle$ $\!=$
$\!\hat{b}_k^\dagger|0\rangle_{b_0}\cdots|0\rangle_{b_N}$}.
Isomorphic mapping can be realized if
\begin{equation}
  \hat{Y}(\omega)
  \sim \hat{P}_{ba}
  := \sum_{k=0}^N|\varphi_k\rangle\;\langle k|.
\label{iso}
\end{equation}
Each state $|\psi\rangle_a$ in $\mathcal{H}_a$ and each operator $\hat{O}_a$
acting on a state in $\mathcal{H}_a$ can then be related to their counterparts
$|\psi\rangle_b$ and $\hat{O}_b$ in $\mathcal{H}_b$, respectively, according to
\begin{equation}
  |\psi\rangle_b = \hat{P}_{ba} |\psi\rangle_a,
\label{1.4a}
\end{equation}
\begin{equation}
  \hat{O}_b = \hat{P}_{ba}\hat{O}_a\hat{P}_{ba}^\dagger,
\label{1.4b}
\end{equation}
and vice versa.

It is well known \cite{ReckZeilinger} that by combining U(2)-beam splitters to
an array one can construct a 2($N$+1)-port which may be described by a
transformation operator $\hat{U}$ obeying
\begin{equation}
  \hat{U}^\dagger\hat{b}_k\hat{U}=\sum_{l=0}^NU_{kl}\hat{b}_l,
\label{ubeamsplitter}
\end{equation}
where the $U_{kl}$ $\!=$ $\!\langle\varphi_k|\hat{U}|\varphi_l\rangle$ can form
any U($N$+1)-group matrix. In $\mathcal{H}_b$, $\hat{U}$ acts therefore as an
arbitrary unitary operator
\begin{equation}
  \hat{U}_b = (\hat{P}_{ba}\hat{P}_{ba}^\dagger)
  \hat{U}(\hat{P}_{ba}\hat{P}_{ba}^\dagger)
  =\sum_{k,l=0}^N|\varphi_k\rangle\langle\varphi_k|
  \hat{U}|\varphi_l\rangle\langle\varphi_l|.
\label{ubeamsplitterb}
\end{equation}
Moreover, measurement-assisted state preparation in $\mathcal{H}_b$
only requires ON/OFF-detectors, which suggests that quantum-state engineering
may be easier in $\mathcal{H}_b$ than in $\mathcal{H}_a$. In order to subject a
state in $\mathcal{H}_a$ to a desired transformation, it could therefore be of
advantage to convert it first into the corresponding state in $\mathcal{H}_b$,
then to transform it there, and eventually the transformed state is converted
back into a state in an isomorphic single-mode space $\mathcal{H}_{a^\prime}$.
%-----------------------------------------------------------------------------
\newpage\section{Implementation of a quantum-state converter}
\label{sec3}
%-----------------------------------------------------------------------------
\subsection{Conditional operation}
\label{sec3.1}
\begin{figure}[htp]
  \centerline{\epsfig{file=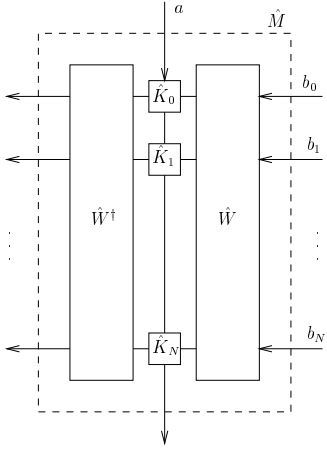,height=6cm}}
  \caption{{\footnotesize
  \label{Fig1}
  Basic scheme of the quantum-state converter $\hat{M}$.
  Two $2(N$ $\!+$ $\!1)$-port beam splitter arrays
  $\hat{W}$ and $\hat{W}^\dagger$ are combined to a $2(N$ $\!+$ $\!1)$-port
  Mach--Zehnder interferometer equipped with
  cross-Kerr couplers $\hat{K}_0,\ldots,\hat{K}_N$.
  }}
\end{figure}

%-----------------------------------------------------------------------------
Let us combine two $2(N$ $\!+$ $\!1)$-port beam splitter arrays $\hat{W}$ and
$\hat{W}^\dagger$ to a $2(N$ $\!+$ $\!1)$-port Mach--Zehnder interferometer and
place inside $N$ $\!+$ $\!1$ cross-Kerr couplers, which realize the
transformations
\begin{equation}
  \hat{K}_k = \mathrm{e}^{\mathrm{i}\kappa_k\hat{b}_k^\dagger\hat{b}_k
  \hat{a}^\dagger\hat{a}}
\end{equation}
as shown in figure~\ref{Fig1}.

The action of the whole device becomes
\begin{eqnarray}
  \hat{M} &=&
  \hat{W}^\dagger\left(\prod_{k=0}^N \hat{K}_{N-k}\right)\hat{W}
  =\hat{W}^\dagger
  \exp\!\left(
  \mathrm{i}\hat{a}^\dagger\hat{a}
  \sum_{k=0}^N\kappa_k\hat{b}_k^\dagger\hat{b}_k
  \right)
  \hat{W}
  \nonumber\\&=&
  \hat{V}^{\hat{a}^\dagger\hat{a}},
\label{intj}
\end{eqnarray}
where
\begin{equation}
  \hat{V} =
  \hat{W}^\dagger
  \exp\left(\mathrm{i}\sum_{k=0}^N\kappa_k\hat{b}_k^\dagger\hat{b}_k\right)
  \hat{W}
\end{equation}
describes a $2(N$ $\!+$ $\!1)$-port beam splitter array
acting in $\mathcal{H}_b$ as a unitary operator
\begin{equation}
  \hat{V}_b = (\hat{P}_{ba}\hat{P}_{ba}^\dagger)
  \hat{V}(\hat{P}_{ba}\hat{P}_{ba}^\dagger)
  =\sum_{k=0}^N \mathrm{e}^{\mathrm{i}\kappa_k}
  \hat{W}^\dagger|\varphi_k\rangle\langle\varphi_k|\hat{W}.
\label{vb}
\end{equation}
The decomposition (\ref{vb}) reveals that the eigenvalues can be
controlled by the strengths $\kappa_k$ of the cross-Kerr interactions and
the eigenbasis by the parameters of the beam splitter array $\hat{W}$.
Thus $\hat{V}_b$ can take the form of any unitary operator in $\mathcal{H}_b$.
Let us choose
\begin{equation}
  \kappa_k = -(N+1)^{-1} 2\pi k,
  \quad 
  \langle\varphi_k|\hat{W}|\varphi_l\rangle
  =(N+1)^{-\frac{1}{2}}\mathrm{e}^{-\mathrm{i}\frac{2\pi kl}{N+1}}.
\end{equation}
In this case, $\hat{V}_b$ takes the form of 
\begin{equation}
  \hat{V}_b=\sum_{k=0}^N|\varphi_{[k+1]}\rangle\langle\varphi_k|,
\label{phaseoperator}
\end{equation}
where $[k+1]=k+1\,\mathrm{mod}\,N+1$.
Note that the single-mode counterpart \mbox{$\hat{V}_a^\dagger$ $\!=$
$\!\hat{P}_{ba}^\dagger\hat{V}_b^\dagger \hat{P}_{ba}$} is the unitary
Pegg--Barnett phase operator \cite{phaseBarnett}.  

\begin{figure}[htp]
  \centerline{\epsfig{file=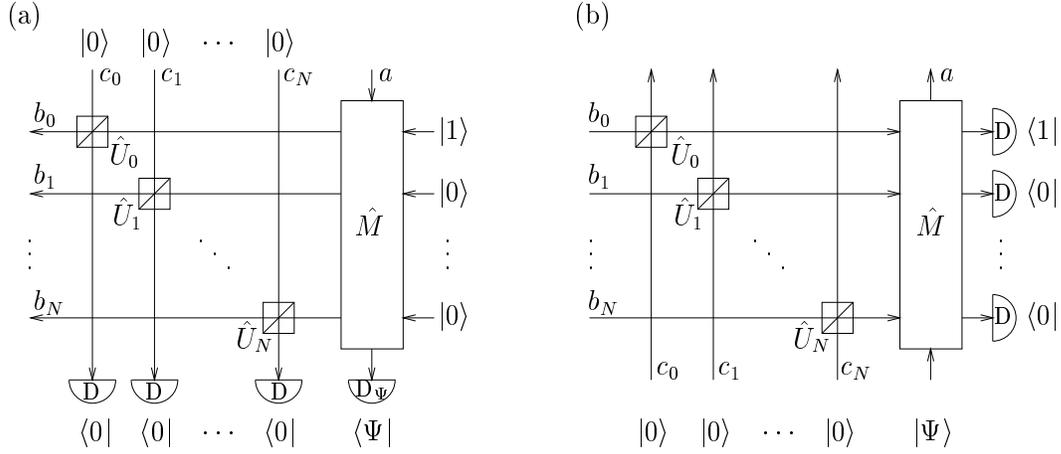,height=6cm}}
  \caption{{\footnotesize
  \label{Fig2}
  Quantum-state converter for (a) realizing the transformation operator
  $\hat{Y}$ given in (\ref{yba}) and (b) the adjoint $\hat{Y}^\dagger$
  ($\hat{M}$, setup according to figure~\ref{Fig1};
  $\hat{U}_k$, beam splitters with transmittances $T_k$;
  D$_\Psi$, device for detection of the state $|\Psi\rangle$;
  D, ON/OFF-detectors for detection of presence or absence of photons).
  }}
\end{figure}

Next, let us specify in more detail the states of the modes and the
measurements to be performed (figure~\ref{Fig2}).

In figure~\ref{Fig2}(a), each of the outgoing $b$-modes of device $\hat{M}$
is coupled to a $c$-mode according to
\begin{equation}
  \hat{U}_k^\dagger\hat{b}_k\hat{U}_k=T_k\hat{b}_k+R_k\hat{c}_k,
  \quad
  |T_k|^2+|R_k|^2=1
\label{tkrk}
\end{equation}
by means of $N$+1 separate beam splitters $\hat{U}_k$. From (\ref{tkrk}) it
follows that
\begin{equation}
  _{c_k}\langle0|\hat{U}_k|0\rangle_{c_k}
  =T_k^{\hat{b}_k^\dagger\hat{b}_k},
\label{tkrkconseq}
\end{equation}
as is shown in the appendix.
The incoming $b$-modes of device $\hat{M}$ are prepared in the state
$|\varphi_0\rangle$, while the incoming $c$-modes are prepared in
vacuum states. If the outgoing $c$-modes are detected in vacuum states
while the outgoing $a$-mode of device $\hat{M}$ is detected in a state
$|\Psi\rangle$, then the combined setup converts, according to
(\ref{yomega}), a $\mathcal{H}_a$-input state of the $a$-mode into a
$\mathcal{H}_b$-output state of the $b$-modes, where the transformation
operator can with (\ref{tkrkconseq}) be written as
\begin{eqnarray}
  \hat{Y}(0_{c_0},...,0_{c_N},\Psi)
  &=&\langle\Psi|
  \left(\prod_{k=0}^N\;_{c_k}\langle0|\hat{U}_k|0\rangle_{c_k}\right)
  \hat{M}|\varphi_0\rangle
  \nonumber\\&=&
  \left(\prod_{k=0}^N T_k^{\hat{b}_k^\dagger\hat{b}_k}
  \right)\langle\Psi|\hat{M}|\varphi_0\rangle.
\label{preyba}
\end{eqnarray}
We assume that none of the Fock expansion coefficients of $|\Psi\rangle$
disappears,
\begin{equation}
  |\langle k|\Psi\rangle|_{\mathrm{min}}:=\min
  \Bigl(|\langle 0|\Psi\rangle|,\ldots,|\langle N|\Psi\rangle|\Bigr)>0,
\end{equation}
and choose 
\begin{equation}
  T_k=\langle\Psi|k\rangle^{-1}|\langle k|\Psi\rangle|_{\mathrm{min}}.
\label{Tk}
\end{equation}
Note that the term $|\langle k|\Psi\rangle|_{\mathrm{min}}$ has been introduced
to obtain \mbox{$|T_k|$ $\!\in$ $\![0,1]$}, thus ensuring the feasibility of
the beam splitter transmittances $T_k$ needed.
Applying (\ref{intj}) and (\ref{phaseoperator}), we see that (\ref{preyba})
takes the form of (\ref{iso}),
\begin{equation}
  \hat{Y}(0_{c_0},...,0_{c_N},\Psi)
  =|\langle k|\Psi\rangle|_{\mathrm{min}}\hat{P}_{ba},
\label{yba}
\end{equation}
and the joint probability (\ref{pomega}) of detecting vacuum states
$|0\rangle_{c_k}$ as well as the state $|\Psi\rangle$ becomes
\begin{eqnarray}
  p(0_{c_0},...,0_{c_N},\Psi)
  &=&\bigl\langle\hat{Y}^\dagger(0_{c_0},...,0_{c_N},\Psi)
  \,\hat{Y}(0_{c_0},...,0_{c_N},\Psi)\bigr\rangle
  \nonumber\\&=&
  |\langle k|\Psi\rangle|_{\mathrm{min}}^2.
\label{pba}
\end{eqnarray}
In figure~\ref{Fig2}(b), the device runs backwards, i.e., the output ports are
used as inputs, replacing detection of a given state with its preparation
and vice versa. These changes are described by replacing (\ref{preyba}) with
its adjoint. That is, if the incoming $a$-mode is prepared in the state
$|\Psi\rangle$ and the incoming $c$-modes are prepared in vacuum states,
while the state
\mbox{$|\varphi_0\rangle$ $\!|0\rangle_{c_0}\cdots|0\rangle_{c_N}$}
is measured by detecting photon presence in the outgoing $b_0$-channel,
then a $\mathcal{H}_b$-input state of the $b$-modes is converted, according to
(\ref{yomega}), into a $\mathcal{H}_a$-output state of the $a$-mode,
where the transformation operator now becomes
\begin{equation}
  \hat{Y}(1_{b_0})
  =\hat{Y}^\dagger(0_{c_0},...,0_{c_N},\Psi)
  =|\langle k|\Psi\rangle|_{\mathrm{min}}\hat{P}_{ba}^\dagger,
\label{yab}
\end{equation}
with the corresponding probability (\ref{pomega}) of detecting photon
presence in the outgoing $b_0$-channel being again
\begin{eqnarray}
  p(1_{b_0})
  &=&\bigl\langle\hat{Y}^\dagger(1_{b_0})\hat{Y}(1_{b_0})\bigr\rangle
  \nonumber\\&=&
  |\langle k|\Psi\rangle|_{\mathrm{min}}^2.
\label{pab}
\end{eqnarray}
Let us now address the question of maximising the probabilities (\ref{pba})
and (\ref{pab}) by a suitable choice of $|\Psi\rangle$. Because of
$\langle\Psi|\Psi\rangle$ $\!=$ $\!\sum_{k=0}^N|\langle k|\Psi\rangle|^2$
$\!=1$, we have
\begin{equation}
  |\langle k|\Psi\rangle|_{\mathrm{min}}^2
  \le(N+1)^{-1}
  =|\langle k|\Phi_\mathrm{P}\rangle|_{\mathrm{min}}^2,
\label{ineq}
\end{equation}
where
\begin{equation}
  |\Phi_\mathrm{P}\rangle:=(N+1)^{-\frac{1}{2}}
  \mathrm{e}^{\mathrm{i}\Phi\hat{a}^\dagger\hat{a}}
  \sum_{k=0}^N|k\rangle
\label{phasestate}
\end{equation}
is a Pegg--Barnett phase state. Comparing (\ref{ineq}) with (\ref{pba}) and
(\ref{pab}), we see that for the state (\ref{phasestate}), the maximal success
probability is achieved. At the same time, (\ref{Tk}) reduces to
$T_k$ $\!=$ $\!\mathrm{exp}(\mathrm{i}k\Phi)$, so that the term in
(\ref{preyba}) depending on the beam splitters $\hat{U}_k$ reads
\begin{equation}
  \prod_{k=0}^N\;_{c_k}\langle0|\hat{U}_k|0\rangle_{c_k}
  =\exp\left(\mathrm{i}\Phi\sum_{k=0}^Nk\hat{b}_k^\dagger\hat{b}_k\right)
  =:\hat{U}_\Phi,
\label{phaskorrekt}
\end{equation}
and the ON/OFF-detectors D in figure~\ref{Fig2}(a) become redundant.
Note that (\ref{phaskorrekt}) acts in $\mathcal{H}_a$ as a unitary operator
\mbox{$\hat{P}_{ba}^\dagger\hat{U}_\Phi \hat{P}_{ba}$ $\!=$
$\!\mathrm{e}^{\mathrm{i}\Phi\hat{a}^\dagger\hat{a}}$}.
%-----------------------------------------------------------------------------
\subsection{Unconditional operation}
\label{sec3.2}
%-----------------------------------------------------------------------------
Let us have a look at figure~\ref{Fig2}(a) and assume that a device D$_\Phi$ is
applied performing a phase measurement such that at each trial some
Pegg--Barnett phase state (\ref{phasestate}) is detected in the outgoing
$a$-mode of device $\hat{M}$. Devices accomplishing detection (or preparation)
of states $|\Phi_\mathrm{P}\rangle$ may be constructed based on the proposals
made in \cite{prepClausen}-\cite{Barnett4}. From (\ref{Tk}) it follows that the
action of the beam splitters $\hat{U}_k$ reduces to (\ref{phaskorrekt}) and the
ON/OFF-detectors D can therefore be removed. The possibility of a
post-measurement implementation of an operator $\hat{U}_{\tilde{\Phi}}$
defined by (\ref{phaskorrekt}) according to the respective phase value
$\tilde{\Phi}$ measured allows us to achieve the desired state transformation
irrespective of $\tilde{\Phi}$. In this way, {\it each} trail results in the
desired state conversion. This can be described according to (\ref{yomega}) if
we use in place of (\ref{yba}) an effective transformation operator
\begin{equation}
  \hat{\Upsilon}=\hat{P}_{ba},
\end{equation}
which includes the effect of post-measurement adaption applied. Accordingly,
the respective success probability (\ref{pomega}) becomes
$\langle\hat{\Upsilon}^\dagger\hat{\Upsilon}\rangle$ $\!=$ $\!1$.

Consider now figure~\ref{Fig2}(b) with the incoming $a$-mode prepared in the
phase state $|\Psi\rangle$ $\!=$ $\!|0_\mathrm{P}\rangle$, so that (\ref{Tk})
gives $T_k$=1 and the beam splitters $\hat{U}_k$ can be removed. Obviously, at
each trial, one of the ON/OFF-detectors D clicks. The probability (\ref{pab})
of a click in channel $b_0$ and with it a successful state conversion becomes
$p(1_{b_0})$ $\!=$ $\!(N+1)^{-1}$. In general however, an ON/OFF-detector in an
arbitrary channel $b_k$ clicks, and instead of (\ref{yab}) we have
\begin{equation}
  \hat{Y}(1_{b_k})
  =\langle\varphi_k|\hat{M}^\dagger|0_\mathrm{P}\rangle
  =(N+1)^{-\frac{1}{2}}\hat{P}_{ba}^\dagger\hat{V}_b^{\dagger\,k}.
\end{equation}
If $k$ $\!\neq$ $\!0$, the ill-transformed output state may now be reconverted
into its counterpart in $\mathcal{H}_b$ [e.g. by feeding it back into the
device $\hat{M}$ using a mirror and running the setup as shown in
figure~\ref{Fig2}(a)]. After subjecting it to a unitary transformation
$\hat{V}_b^k$ (implemented by a beam splitter array in the $b$-channels), the
original input state is reobtained. By feeding it again back into the setup
figure~\ref{Fig2}(b) (e.g. by using another mirror in the $b$-channels), we can
start a new trial of state conversion. This procedure of bouncing the signal
forward and backward is continued until eventually the ON/OFF-detector in
channel $b_0$ clicks. The average number of trials needed until this happens is
$\bar{n}$ $\!=$ $\!p(1_{b_0})^{-1}$ $\!=$ $\!N$+1. In this way, {\it each}
$\mathcal{H}_b$-input state of the $b$-modes is sooner or later properly
converted into its $\mathcal{H}_a$-counterpart of the $a$-mode. This can again
be described according to (\ref{yomega}) if we use in place of
(\ref{yab}) an effective transformation operator
\begin{equation}
  \hat{\Upsilon}=\hat{P}_{ba}^\dagger,
\end{equation}
which includes all the procedures applied.
Again, the respective success probability (\ref{pomega}) becomes
$\langle\hat{\Upsilon}^\dagger\hat{\Upsilon}\rangle$ $\!=$ $\!1$.
In summary, we see that the setups figure~\ref{Fig2} allow in principle a
reversible conversion between states in the single-mode space $\mathcal{H}_a$
and their counterparts in the multi-mode space $\mathcal{H}_b$.
%-----------------------------------------------------------------------------
\newpage\section{Application to quantum state engineering}
\label{sec4}
%-----------------------------------------------------------------------------
\subsection{Conditional operation}
\label{sec4.1}
\begin{figure}[htp]
  \centerline{\epsfig{file=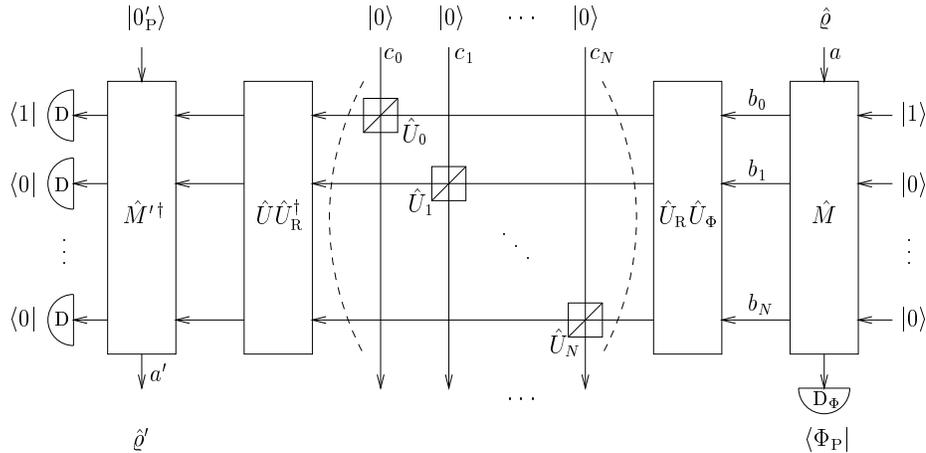,height=6cm}}
  \caption{{\footnotesize
  \label{Fig3}
  Combination of two quantum-state converters (figure~\ref{Fig2}) with
  beam splitter arrays for realizing arbitrary transformations of single-mode
  quantum states \mbox{$\hat{\varrho}$ $\!\to$ $\!\hat{\varrho}'$}
  and arbitrary single-mode quantum state measurement (with the incoming
  $a^\prime$-mode of the left-hand converter prepared in the vacuum state
  instead of $|0_\mathrm{P}^\prime\rangle$; see section~\ref{sec5}).
  The elements in the dashed parentheses can be omitted if attention is
  restricted to unitary state engineering or orthogonal basis measurements.
  }}
\end{figure}

%-----------------------------------------------------------------------------
Let us now consider the problem of realizing an arbitrary unitary or
non-unitary transformation of a single-mode state $\hat{\varrho}$ into another
single-mode state $\hat{\varrho}^\prime$ according to (\ref{yomega}).
For this purpose, we combine the two state converters in figure~\ref{Fig2} with
two \mbox{$2(N$ $\!+$ $\!1)$}-port beam splitter arrays $\hat{U}_\mathrm{R}$
and $\hat{U}\hat{U}_\mathrm{R}^\dagger$ and \mbox{$N$ $\!+$ $\!1$} beam
splitters $\hat{U}_k$, as shown in figure~\ref{Fig3}. Since detection of a
state (\ref{phasestate}) is assumed in the right-hand converter, the action of
the separate beam splitters (and ON/OFF-detectors) in figure~\ref{Fig2}(a)
reduces to (\ref{phaskorrekt}), and since preparation of the state
(\ref{phasestate}) with $\Phi$ $\!=$ $\!0$ is assumed in the left-hand
converter, the separate beam splitters in figure~\ref{Fig2}(b) are left out.

The detection of photon presence in the outgoing $b_0$-channel of the left-hand
converter is equivalent to the detection of the state
\mbox{$|\varphi_0\rangle$ $\!|0\rangle_{c_0}\cdots|0\rangle_{c_N}$}, so that,
on measuring the state $|\Phi_\mathrm{P}\rangle$ in the single-mode output
channel of the right-hand converter and using the state
$|0_\mathrm{P}\rangle$ as the single-mode input of the left-hand converter,
the single-mode input state $\hat{\varrho}$ is related to the single-mode
output state $\hat{\varrho}^\prime$ according to (\ref{yomega}), where
\begin{eqnarray}
  \hat{Y}(1_{b_0},\Phi)
  &=&\langle\varphi_0|\langle\Phi_\mathrm{P}|
  \hat{M}^{\prime\,\dagger}\hat{A}\hat{U}_\Phi\hat{M}
  |0_\mathrm{P}^\prime\rangle|\varphi_0\rangle
  \nonumber\\&=&
  (N+1)^{-1}\hat{P}_{ba^\prime}^\dagger\hat{A}_b\hat{P}_{ba} =
  (N+1)^{-1}\hat{P}_{ba^\prime}^\dagger\hat{P}_{ba}\hat{A}_a.
\label{ya'a}
\end{eqnarray}
Here, we have set $\hat{A}$ $\!=$ $\!\hat{U}\hat{R}$ with
\begin{equation}
  \hat{R}
  =\hat{U}_\mathrm{R}^\dagger
  \left(\prod_{k=0}^N\;_{c_k}\langle0|\hat{U}_k|0\rangle_{c_k}\right)
  \hat{U}_\mathrm{R}
  =\hat{U}_\mathrm{R}^\dagger
  \left(\prod_{k=0}^NT_k^{\hat{b}_k^\dagger\hat{b}_k}\right)
  \hat{U}_\mathrm{R}
\end{equation}
acting in $\mathcal{H}_b$ as an operator
\begin{equation}
  \hat{R}_b
  =(\hat{P}_{ba}\hat{P}_{ba}^\dagger)
  \hat{R}(\hat{P}_{ba}\hat{P}_{ba}^\dagger)
  =\sum_{k=0}^N T_k
  \hat{U}_\mathrm{R}^\dagger|\varphi_k\rangle
  \langle\varphi_k|\hat{U}_\mathrm{R}.
\label{r}
\end{equation}
In the first line of (\ref{ya'a}), we can replace $\hat{A}$ with
\begin{equation}
  \hat{A}_b=\hat{U}_b\hat{R}_b
\label{italicpolar}
\end{equation}
and introduce
\mbox{$\hat{A}_a$ $\!=$ $\!\hat{P}_{ba}^\dagger\hat{A}_b\hat{P}_{ba}$},
which, together with (\ref{yba}) and (\ref{yab}), gives the second line of
(\ref{ya'a}). The decomposition (\ref{r}) reveals that the eigenbasis of
$\hat{R}_b$ can be controlled via the beam splitter array $\hat{U}_\mathrm{R}$,
while the eigenvalues can be varied by the transmittances $T_k$ of the $N$+1
beam splitters $\hat{U}_k$ in the dashed-line parentheses. In particular, by
choosing the $T_k$ to be positive real, \mbox{$T_k$ $\!\in$ $\![0,1]$}, with
the extra condition \mbox{$\sum_{k=0}^NT_k$ $\!=$ $\!1$}, $\hat{R}_b$ takes the
form of any desired density-type operator.
Assume now that there is the task of implementing the state transformation
(\ref{yomega}) associated with a given operator
$\hat{\mathrm{A}}\equiv\hat{\mathrm{A}}_b$ acting in $\mathcal{H}_b$.
Its polar decomposition $\hat{\mathrm{A}}$ $\!=$ 
$\!\hat{\mathrm{U}}(\hat{\mathrm{A}}^\dagger\hat{\mathrm{A}})^\frac{1}{2}$
(the unitary operator $\hat{\mathrm{U}}$ is unique if $\hat{\mathrm{A}}$ has an
inverse) can for \mbox{$\hat{\mathrm{A}}$ $\neq \hat{0}$} be written as
\begin{equation}
  \hat{\mathrm{A}}
  =\mathrm{Tr}(\hat{\mathrm{A}}^\dagger\hat{\mathrm{A}})^\frac{1}{2}
  (\mathrm{Det}\hat{\mathrm{U}})^\frac{1}{N+1}\;
  \hat{\mathrm{A}}_\mathrm{norm},
\end{equation}
where $\mathrm{Det}\hat{\mathrm{U}}$ $\!=$
$\!\exp(\mathrm{Tr}\ln\hat{\mathrm{U}}$) and 
\begin{equation}
  \hat{\mathrm{A}}_\mathrm{norm}
  =\frac{\hat{\mathrm{U}}}{(\mathrm{Det}\hat{\mathrm{U}})^\frac{1}{N+1}}\;
  \frac{(\hat{\mathrm{A}}^\dagger\hat{\mathrm{A}})^\frac{1}{2}}
  {\mathrm{Tr}(\hat{\mathrm{A}}^\dagger\hat{\mathrm{A}})^\frac{1}{2}}
\label{romanpolar}
\end{equation}
is the product of an SU($N$+1)- and a density-type operator. By identifying
(\ref{romanpolar}) with (\ref{italicpolar}), we see that any desired
state transformation (\ref{yomega}) defined by a given operator
$\hat{\mathrm{A}}$ can be implemented. $\hat{\mathrm{A}}_\mathrm{norm}$ here
serves as normalized representant of the class of all operators differing by
multiplication with a c-number factor but giving rise to the same state
transformation (\ref{yomega}).

The factor $\mathrm{Tr}(\hat{\mathrm{A}}^\dagger\hat{\mathrm{A}})^\frac{1}{2}$
however alters the success probability (\ref{pomega}). Since we are
not restricted to the extra condition \mbox{$\sum_{k=0}^NT_k$ $\!=$ $\!1$}, we
may therefore choose the \mbox{$T_k$ $\!\in$ $\![0,1]$} in order to maximize
this probability, i.e., the joint probability
\begin{eqnarray}
  p(1_{b_0},\Phi)
  &=&\bigl\langle\hat{Y}^\dagger(1_{b_0},\Phi)\hat{Y}(1_{b_0},\Phi)\bigr\rangle
  \nonumber\\&=&
  (N+1)^{-2}\bigl\langle\hat{R}_a^\dagger\hat{R}_a\bigr\rangle
\label{pa'a}
\end{eqnarray}
(\mbox{$\hat{R}_a$ $\!=$ $\!\hat{P}_{ba}^\dagger\hat{R}_b\hat{P}_{ba}$})
of detecting photon presence in the outgoing $b_0$-channel and measuring the
state $|\Phi_\mathrm{P}\rangle$.
We see that $p(1_{b_0},\Phi)$ depends in general on the input state
$\hat{\varrho}$ and may even identically vanish for certain inputs.
(The case \mbox{$\hat{A}_b$ $\!=$ $\!\hat{0}$} corresponding to
\mbox{$T_k$ $\!=$ $\!0$} for which this happens for all input states has been
excluded.) Two types of state transformations are of special interest:
\begin{itemize}
\item[(i)]{\it Unitary transformations.}

They are obtained by choosing the values of the $T_k$ to be identical, $T_k=T$,
so that (\ref{italicpolar}) reduces to
$\hat{A}_b$ $\!=$ $\!\hat{U}_bT$ $\!\sim$ $\!\hat{U}_b$, while the
success probability (\ref{pa'a}) becomes independent of the input state,
$p(1_{b_0},\Phi)$ $\!=$ \mbox{$\!T^2(N$ $\!+$ $\!1)^{-2}$}.
Of particular interest is the case \mbox{$T$ $\!=$ $\!1$}, allowing removal of
the $\hat{U}_k$, thus simplifying the setup while the same time the success
probability is maximized.
\item[(ii)]{\it Projective transformations.}

They are obtained by choosing
\mbox{$\hat{U}_b$ $\!=$ $\!\hat{P}_{ba}\hat{P}_{ba}^\dagger$},
so that (\ref{italicpolar}) reduces to a positive operator,
$\hat{A}_b$ $\!=$ $\!\hat{R}_b$, and confining the allowed values of the $T_k$
to $0$ and some given $T$ $\!\in$ $\![0,1]$, so that
$\hat{A}_b$ $\!\sim$ $\!\hat{A}_b^2$ $\!\sim$ $\!\hat{A}_b^\dagger$.
Of particular interest is the case
\mbox{$T_k$ $\!=$ $\!\delta_{kl}$}, leading in (\ref{ya'a}) to 
\begin{equation}
  \hat{A}_a = \hat{U}_{R,a}^\dagger|l\rangle\langle l|\hat{U}_{R,a}
\label{operatorprojective}
\end{equation}
(\mbox{$\hat{U}_{R,a}$ $\!=$
$\!\hat{P}_{ba}^\dagger\hat{U}_\mathrm{R}\hat{P}_{ba}$}).
The output state $\hat{\varrho}^\prime$ then becomes independent of the input
state $\hat{\varrho}$, since in this case a projection onto the state
$\hat{U}_{R,a}^\dagger|l\rangle$ is performed, and the success probability
(\ref{pa'a}) reads
\begin{equation}
  p(1_{b_0},\Phi)=(N+1)^{-2}\langle l|\hat{U}_{R,a}\hat{\varrho}
  \hat{U}_{R,a}^\dagger|l\rangle.
\label{probabilityprojective}
\end{equation}
\end{itemize}
%-----------------------------------------------------------------------------
\subsection{Unconditional operation}
\label{sec4.2}
%-----------------------------------------------------------------------------
So far, attention has been limited to the event of detecting photon presence in
the outgoing $b_0$-channel of the left-hand converter as well as a detecting
in the right-hand converter a phase parameter $\Phi$ given by the beam splitter
array $\hat{U}_\Phi$. Alternatively, we may apply an $\hat{U}_{\tilde{\Phi}}$
depending on the respective value $\tilde{\Phi}$ measured and, if photon
presence has been detected in one of the $b$-channels of the left-hand
converter, apply the procedure described in section (\ref{sec3.2}) to obtain
unconditional convertion into $\mathcal{H}_{a^\prime}$. These additional
procedures are taken into account by using in place of (\ref{ya'a}) an
effective transformation operator
\begin{eqnarray}
  \hat{\Upsilon}(1_b)
  &=&\hat{P}_{ba^\prime}^\dagger\hat{A}_b\hat{P}_{ba}
  =\hat{P}_{ba^\prime}^\dagger\hat{P}_{ba}\hat{A}_a
  \nonumber\\&=&
  (N+1)\;\hat{Y}(1_{b_0},\Phi),
\label{yeffa'a}
\end{eqnarray}
and in place of (\ref{pa'a}) we obtain
\begin{eqnarray}
  p(1_b)
  &=&\bigl\langle\hat{\Upsilon}^\dagger(1_b)\hat{\Upsilon}(1_b)\bigr\rangle
  =\bigl\langle\hat{R}_a^\dagger\hat{R}_a\bigr\rangle
  \nonumber\\&=&
  (N+1)^2\;p(1_{b_0},\Phi)
\label{peffa'a}
\end{eqnarray}
as the probability of detecting photon presence in one of the $b$-channels of
the left-hand converter. In particular, if the beam splitters in dashed
parentheses in figure~\ref{Fig3} are removed, $T_k=1$, then at each trial,
one of the ON/OFF-detectors D clicks, $p(1_b)=1$. In this way, any unitary
transformation can be realized unconditionally.
%-----------------------------------------------------------------------------
\section{Application to quantum-state measurement}
\label{sec5}
%-----------------------------------------------------------------------------
\subsection{Conditional operation}
\label{sec5.1}
%-----------------------------------------------------------------------------
We now consider the scheme in figure~\ref{Fig3} with the incoming
$a^\prime$-mode of the left-hand converter prepared in the vacuum state
$|0^\prime\rangle$ instead of $|0_\mathrm{P}^\prime\rangle$. In the event of
registering photon presence in the channel $b_k$ and detecting the state
$|\Phi_\mathrm{P}\rangle$ in the single-mode output channel of the right-hand
converter, (\ref{ya'a}) is replaced with
\begin{eqnarray}
  \hat{Y}_0(1_{b_k},\Phi)
  &=&\langle\varphi_k|\langle\Phi_\mathrm{P}|
  \hat{M}^{\prime\,\dagger}\hat{A}\hat{U}_\Phi\hat{M}
  |0^\prime\rangle|\varphi_0\rangle
  \nonumber\\&=&
  (N+1)^{-\frac{1}{2}}|0^\prime\rangle\langle k|\hat{A}_a,
\end{eqnarray}
and the corresponding joint probability of registering photon presence in
channel $b_k$ and the state $|\Phi_\mathrm{P}\rangle$ in the right-hand
converter is given by
\begin{eqnarray}
  p(1_{b_k},\Phi)
  &=&\bigl\langle\hat{Y}_0^\dagger(1_{b_k},\Phi)
  \hat{Y}_0(1_{b_k},\Phi)\bigr\rangle
  \nonumber\\&=&
  (N+1)^{-1}\langle k|\hat{A}_a\hat{\varrho}\hat{A}_a^\dagger|k\rangle.
\label{pk}
\end{eqnarray}
In this way, on recalling that $\hat{A}_a^\dagger$ may be defined by its action
on the basis states $|k\rangle$, the overlaps of a state $\hat{\varrho}$ with
the members of an arbitrary set of states
$\langle k|\hat{A}_a\hat{A}_a^\dagger|k\rangle^{-\frac{1}{2}}
\hat{A}_a^\dagger|k\rangle$,
\mbox{$k$ $\!=$ $\!0,\ldots,N$}, can be measured.

Of special interest is again the case when \mbox{$T_k$ $\!=$ $\!1$} for all $k$
(figure~\ref{Fig3} without the beam splitters in the dashed parentheses), so
that \mbox{$\hat{A}_a$ $\!=$ $\!\hat{P}_{ba}^\dagger\hat{U}_b\hat{P}_{ba}$
$\!=$ $\!\hat{U}_a$}. The probability of detecting photon presence in the
channel $b_k$ conditioned on the measurement of $|\Phi_\mathrm{P}\rangle$ thus
becomes
\begin{equation}
  p(1_{b_k}|\Phi)=\frac{p(1_{b_k},\Phi)}{p(\Phi)}
  =\langle k|\hat{U}_a\hat{\varrho}\hat{U}_a^\dagger|k\rangle,
\label{pkcond}
\end{equation}
where $p(\Phi)$ follows from (\ref{pba}).
In this way, the overlaps of the input state $\hat{\varrho}$ with states
$\hat{U}_a^\dagger|k\rangle$, \mbox{$k$ $\!=$ $\!0,\ldots,N$}, forming an
orthonormal basis controlled by $\hat{U}$ can be measured.

Measurements in two orthonormal bases are necessary to determine the
expectation value of a given operator
$\hat{\mathrm{Z}}\equiv\hat{\mathrm{Z}}_a$ acting in $\mathcal{H}_a$. This is
seen from its Cartesian decomposition
$\hat{\mathrm{Z}}$ $\!=$ $\!\hat{\mathrm{Z}}_{\mathrm{Re}}$ $\!+$
$\!\mathrm{i}\hat{\mathrm{Z}}_{\mathrm{Im}}$ into the Hermitian operators 
$\hat{\mathrm{Z}}_{\mathrm{Re}}$ $\!=$
$\!(\hat{\mathrm{Z}}+\hat{\mathrm{Z}}^\dagger)/2$ and
$\hat{\mathrm{Z}}_{\mathrm{Im}}$ $\!=$
$\!(\hat{\mathrm{Z}}-\hat{\mathrm{Z}}^\dagger)/2\mathrm{i}$.
Inserting the spectral decomposition of $\hat{\mathrm{Z}}_{\mathrm{Re}}$ and
$\hat{\mathrm{Z}}_{\mathrm{Im}}$, we obtain
\begin{equation}
  \bigl\langle\hat{\mathrm{Z}}\bigr\rangle=\sum_{j=0}^1\sum_{k=0}^N
  \mathrm{i}^j\lambda_{jk}
  \langle k|\hat{U}_{a,j}\hat{\varrho}\hat{U}_{a,j}^\dagger|k\rangle,
\end{equation}
where the $\lambda_{jk}$ and $\hat{U}_{a,j}^\dagger|k\rangle$ are the
eigenvalues and eigenstates of $\hat{\mathrm{Z}}_{\mathrm{Re}}$ ($j$=0) and
$\hat{\mathrm{Z}}_{\mathrm{Im}}$ ($j$=1), respectively. 
An example is $\hat{\mathrm{Z}}$ $\!=$ $\!|n\rangle\langle m|$, for which
$\lambda_{jk}=(\delta_{kn}-\delta_{km})/2$ so that 
$\langle\hat{\mathrm{Z}}\rangle$ $\!=$ $\!\langle m|\hat{\varrho}|n\rangle$
is determined by measurements in the channels $b_n$ and $b_m$ alone,
which are coupled by a symmetric beam splitter defined by
$\langle\varphi_n|\hat{U}_j^\dagger|\varphi_n\rangle$ $\!=$ 
$\langle\varphi_m|\hat{U}_j^\dagger|\varphi_m\rangle$ $\!=$ 
$\mathrm{i}^{-j}\langle\varphi_m|\hat{U}_j^\dagger|\varphi_n\rangle$ $\!=$ 
$-\mathrm{i}^{-j}\langle\varphi_n|\hat{U}_j^\dagger|\varphi_m\rangle^\ast$
$\!=$ $\!2^{-\frac{1}{2}}$. Repeating these measurements for different $m$
and $n$ allows us to reconstruct the input state $\hat{\varrho}$ from its
matrix elements $\langle m|\hat{\varrho}|n\rangle$ in the Fock basis.

The task of determining an unknown input state $\hat{\varrho}$ may also be
accomplished experimentally, without a subsequent reconstruction from measured
data. This is seen from (\ref{pkcond}). Under variation of $\hat{U}$,
$p(1_{b_k}|\Phi)$ becomes extremal iff $\hat{U}_a^\dagger|k\rangle$ is an
eigenstate of $\hat{\varrho}$ \cite{SchubWeb}. On the other hand,
$p(1_{b_k}|\Phi)$ cannot be greater than the greatest eigenvalue of
$\hat{\varrho}$. We implement $\hat{U}$ according to
\begin{equation}
  \hat{U}=\hat{U}_{(N-1)N}\cdots\hat{U}_{1\ldots N}\hat{U}_{0\ldots N}
\label{overall}
\end{equation}
as a chain of $2(N$ $\!-$ $\!k$ $\!+$ $\!1)$-port beam splitter arrays
$\hat{U}_{k\ldots N}$ coupling the channels $b_k,\ldots,b_N$ to each other, and
start by tuning $\hat{U}_{0\ldots N}$ until the detector signal
$p(1_{b_0}|\Phi)$ obtained in channel $b_0$ is maximal.
[The term detector signal in a given channel $b_k$ is here used for the
relative frequency of clicks obtained with the ON/OFF-detector in this channel
given the detection of $|\Phi_\mathrm{P}\rangle$ under repetition of the
experiment which approximates the probability $p(1_{b_k}|\Phi)$.] With regard
to the operator $\hat{U}$ obtained in this way, the eigenvalue equation
\begin{equation}
  \hat{\varrho}\;\hat{U}_a^\dagger|k\rangle
  = p(1_{b_k}|\Phi)\;\hat{U}_a^\dagger|k\rangle
\label{ewp}
\end{equation}
holds for $k$=0. We now keep $\hat{U}_{0\ldots N}$ fixed and tune
$\hat{U}_{1\ldots N}$ until the detector signal $p(1_{b_1}|\Phi)$ obtained in
channel $b_1$ is maximal. With this new resulting tuned overall operator
(\ref{overall}), the eigenvalue equation (\ref{ewp}) holds for $k$=0,1. This
procedure is continued until all beam splitter arrays $\hat{U}_{k\ldots N}$,
$k=0,\ldots,N-1$, are tuned and (\ref{ewp}) holds for all $k$. With the
resulting operator $\hat{U}$, we can now write the input state as
\begin{equation}
  \hat{\varrho}=\sum_{k=0}^N p(1_{b_k}|\Phi)
  \;\hat{U}_a^\dagger|k\rangle\langle k|\hat{U}_a,
\label{inputstate}
\end{equation}
where the $p(1_{b_k}|\Phi)$ are the respective maximized detector signals
obtained in channel $b_k$. These are the eigenvalues of $\hat{\varrho}$ in
descending order. Note that in this way, the input state $\hat{\varrho}$ has
been diagonalized experimentally in the basis defined by the channels, because
the multi-mode state entering the ON/OFF-detectors in figure~\ref{Fig3} reads
$\hat{U}\hat{\varrho}_b\hat{U}^\dagger$ $\!=$
$\!\sum_{k=0}^N p(1_{b_k}|\Phi)|\varphi_k\rangle\langle\varphi_k|$,
where \mbox{$\hat{\varrho}_b$ $\!=$
$\!\hat{P}_{ba}\hat{\varrho}\hat{P}_{ba}^\dagger$} is the converted state
leaving the multi-mode output port of the right-hand converter, i.e., exiting
the array $\hat{U}_\Phi$.

On the other hand, the $p(1_{b_k}|\Phi)$ cannot be smaller than the smallest
eigenvalue of $\hat{\varrho}$, so that, instead of maximizing, we may equally
well successively minimize the detector signals $p(1_{b_k}|\Phi)$. With these
minimized detector signals and the operator $\hat{U}$ resulting from this
procedure, we can write the input state again in the form of
(\ref{inputstate}), where the $p(1_{b_k}|\Phi)$ are now the eigenvalues of
$\hat{\varrho}$ in ascending order. However, the maximizing procedure has the
advantage that the maximized detector signal $p(1_{b_0}|\Phi)$ obtained after
the first step gives us information about the purity of the input state
$\hat{\varrho}$. If for the maximal value $p(1_{b_0}|\Phi)$ $\!=$ $\!1$ holds,
then the input is pure, $\hat{\varrho}$ $\!=$ $\!\hat{\varrho}^2$, and if
$p(1_{b_0}|\Phi)$ comes close to one we may stop the procedure after few steps
if only approximate knowledge about $\hat{\varrho}$ is desired.

There is also the possibility of a quantum non-demolition measurement of pure
and almost pure states. To see this, consider the situation as described by
(\ref{operatorprojective}) and (\ref{probabilityprojective}). After tuning
$\hat{U}_\mathrm{R}$ such that the probability (\ref{probabilityprojective})
has become maximal, (\ref{operatorprojective}) realizes a ``purification'' of
the single-mode input state $\hat{\varrho}$ while the optimized probability
(\ref{probabilityprojective}) tells us the overlap (fidelity) between
(the unknown) input state $\hat{\varrho}$ and the (known) output state
$\hat{\varrho}^\prime$ $\!\equiv$ $\hat{A}_{a^\prime}$ $\!=$
$\!(\hat{P}_{ba^\prime}^\dagger\hat{P}_{ba})\hat{A}_a
(\hat{P}_{ba^\prime}^\dagger\hat{P}_{ba})^\dagger$.
%-----------------------------------------------------------------------------
\subsection{Unconditional operation}
\label{sec5.2}
%-----------------------------------------------------------------------------
Again, we may apply an $\hat{U}_{\tilde{\Phi}}$ depending on the respective
value $\tilde{\Phi}$ measured, instead of considering only such events in which
some fixed phase value has been detected. In place of (\ref{pk}) we therefore
now have
\begin{eqnarray}
  p(1_{b_k})&=&\langle k|\hat{A}_a\hat{\varrho}\hat{A}_a^\dagger|k\rangle
  \nonumber\\&=&
  (N+1)\;p(1_{b_k},\Phi).
\end{eqnarray}
In this way, the measurement time can be reduced by the factor $N$+1.
%-----------------------------------------------------------------------------
\newpage\section{Application to quantum telemanipulation}
\label{sec6}
%-----------------------------------------------------------------------------
\subsection{Conditional operation}
\label{sec6.1}
\begin{figure}[htp]
  \centerline{\epsfig{file=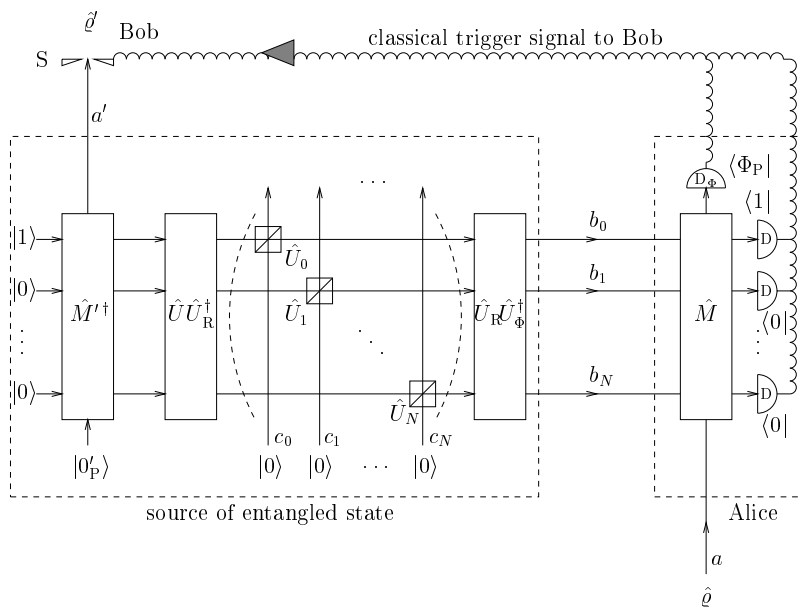,height=8cm}}
  \caption{{\footnotesize
  \label{Fig4}
  Scheme for quantum telemanipulation. 
  The input and output ports of the optical components in figure~\ref{Fig3}
  are interchanged.
  Bob opens the shutter S only after having received confirmation
  from Alice that proper state detection has been realized.
  }}
\end{figure}

%-----------------------------------------------------------------------------
Consider the setup figure~\ref{Fig3}. The initial state $\hat{\varrho}$ of the
incoming $a$-mode of the right-hand converter is destroyed, whereas an
engineered state $\hat{\varrho}^\prime$ of the spatially separated outgoing
$a^\prime$-mode emerges at the left-hand converter. This switch from mode $a$
to mode $a^\prime$ is described in (\ref{ya'a}) by the operator
$\hat{P}_{ba^\prime}^\dagger\hat{P}_{ba}$, which represents an isomorphism
between the single-mode spaces $\mathcal{H}_a$ and $\mathcal{H}_{a^\prime}$.
To operate the scheme it is not essential at which side the single-photon state
$|1\rangle$ is prepared and at which it is detected. This is illustrated in
figure~\ref{Fig4}, in which the output ports of the optical components in
figure~\ref{Fig3} are used as input ports and vice versa. Moreover,
$\hat{U}_\Phi$ is replaced with $\hat{U}_\Phi^\dagger$.

Instead of (\ref{ya'a}), we now have
\begin{eqnarray}
  \hat{Y}_{\mathrm{tel}}(1_{b_0},\Phi)
  &=&\langle\varphi_0|\langle\Phi_\mathrm{P}|
  \hat{M}^\dagger\hat{U}_\Phi\hat{A}^\dagger\hat{M}^\prime
  |0_\mathrm{P}^\prime\rangle|\varphi_0\rangle
  \nonumber\\&=&
  (N+1)^{-1}\hat{P}_{ba^\prime}^\dagger\hat{A}_b^\ast\hat{P}_{ba}=
  (N+1)^{-1}\hat{P}_{ba^\prime}^\dagger\hat{P}_{ba}\hat{A}_a^\ast
  \nonumber\\&=&
  \hat{Y}^\ast(1_{b_0},\Phi)
\label{ytela'a}
\end{eqnarray}
(\mbox{$\hat{A}_a^\ast$ $\!=$
$\!\hat{P}_{ba}^\dagger\hat{A}_b^\ast\hat{P}_{ba}$}).
Comparing (\ref{ytela'a}) with (\ref{ya'a}), we see that the only change that
must be made when going from figure~\ref{Fig3} to figure~\ref{Fig4} is
replacing $\hat{A}_b$ with its complex conjugate $\hat{A}_b^\ast$.
However, the setup in figure~\ref{Fig4} performs, in contrast to that in
figure~\ref{Fig3}, the state engineering as a conditional teleportation.
This is seen by distinguishing between Alice, Bob, and the
entangled state source as shown in the figure. Bob operates an optical shutter
S letting his signal prepared in the output state $\hat{\varrho}^\prime$ pass
only if he has received a classical trigger bit from Alice confirming the
detection of the states $|\Phi_\mathrm{P}\rangle$ and $|\varphi_0\rangle$ in
the respective output channels of her state converter. The beam splitter arrays
and the left-hand converter in figure~\ref{Fig4} can be regarded as forming the
source of the entangled state with a multi-mode output to Alice and a
single-mode output to Bob. Since Bob's state $\hat{\varrho}^\prime$ can differ
from Alice's state $\hat{\varrho}$ by an arbitrarily chosen (unitary or
non-unitary) transformation, the scheme combines teleportation and state
engineering. In this sense, we may adopt the term ``telemanipulation''.
Clearly, when the source of the entangled state only consists of
the left converter, so that $\hat{A}_a^\ast=\hat{P}_{ba}^\dagger\hat{P}_{ba}$,
(\ref{ytela'a}) reduces to the mode switch operator mentioned in the beginning,
$\hat{Y}_{\mathrm{tel}}(1_{b_0},\Phi)$ $\!\sim$
$\!\hat{P}_{ba^\prime}^\dagger\hat{P}_{ba}$, and Bob's state becomes a copy of
Alice's state. 
 
In order to investigate the differences between the schemes figure~\ref{Fig3}
and figure~\ref{Fig4} we now remove the device $\mathrm{D}_\Phi$ in both setups
and consider the reduced single-mode states $\hat{\varrho}_{\mathrm{red}}$ and
$\hat{\varrho}_{\mathrm{red}}^\prime$ which are obtained at the $a$- and
$a^\prime$-output ports in that case. We start with figure~\ref{Fig3}.
At the single-mode output of the right-hand converter we obtain
\begin{eqnarray}
  \hat{\varrho}_{\mathrm{red}}
  &:=&\mathrm{Tr}_{\mathcal{H}_b}
  \left(\hat{M}|\varphi_0\rangle\hat{\varrho}
  \langle\varphi_0|\hat{M}^\dagger\right)
  \nonumber\\&=&
  \sum_{k=0}^N|k\rangle\langle k|\hat{\varrho}|k\rangle\langle k|
\label{rho0fig3}
\end{eqnarray}
and at the single-mode output of the left-hand converter ($\hat{X}_3$ $\!:=$
$\!\hat{M}^{\prime\,\dagger}\hat{A}\hat{U}_\Phi\hat{M}$)

\begin{eqnarray}
  \hat{\varrho}_{\mathrm{red}}^\prime
  &:=&\frac{1}{p(1_{b_0})}\mathrm{Tr}_{\mathcal{H}_a}
  \left(\langle\varphi_0|\hat{X}_3
  |\varphi_0\rangle|0_\mathrm{P}^\prime\rangle\hat{\varrho}
  \langle0_\mathrm{P}^\prime|\langle\varphi_0|
  \hat{X}_3^\dagger|\varphi_0\rangle\right)
  \nonumber\\&=&
  \frac{1}{p(1_{b_0})}
  \hat{Y}_{\mathrm{red}}(1_{b_0})
  \;\hat{\varrho}_{\mathrm{red}}\;
  \hat{Y}_{\mathrm{red}}^\dagger(1_{b_0}),
\label{rho1fig3}
\end{eqnarray}
where $p(1_{b_0})$ $\!=$
$\!\mathrm{Tr}_{\mathcal{H}_a}[\hat{\varrho}_{\mathrm{red}}
\hat{Y}_{\mathrm{red}}^\dagger(1_{b_0})\hat{Y}_{\mathrm{red}}(1_{b_0})]$ and
$\hat{Y}_{\mathrm{red}}(1_{b_0})$ $\!=$
$\!(N+1)^{\frac{1}{2}}\hat{Y}(1_{b_0},\Phi)$ with $\hat{Y}(1_{b_0},\Phi)$ given
in (\ref{ya'a}). We see that both states depend on the input state
$\hat{\varrho}$. If we remove the components between the converters, so that
$\hat{A}_a=\hat{P}_{ba}^\dagger\hat{P}_{ba}$, then (\ref{rho1fig3}) coincides
with (\ref{rho0fig3}), $\hat{\varrho}_{\mathrm{red}}^\prime$ $\!=$
$\!(\hat{P}_{ba^\prime}^\dagger\hat{P}_{ba})\hat{\varrho}_{\mathrm{red}}
(\hat{P}_{ba^\prime}^\dagger\hat{P}_{ba})^\dagger$.
If in particular the input state is a mixture of Fock states,
$\hat{\varrho}=\varrho(\hat{a}^\dagger\hat{a})$, we obtain additionally
$\hat{\varrho}_{\mathrm{red}}$ $\!=$ $\!\hat{\varrho}$, so that
(\ref{rho0fig3}) and (\ref{rho1fig3}) are two identical copies (``clones'') of
the input state $\hat{\varrho}$ in this case.

We now turn to figure~\ref{Fig4} which gives at the single-mode output
of the right-hand converter ($\hat{X}_4$ $\!:=$
$\!\hat{M}^{\prime\,\dagger}\hat{A}\hat{U}_\Phi^\dagger\hat{M}$)
\begin{eqnarray}
  \hat{\varrho}_{\mathrm{red}}
  &:=&\frac{1}{p(1_{b_0})}\mathrm{Tr}_{\mathcal{H}_{a^\prime}}
  \left(\langle\varphi_0|\hat{X}_4^\dagger
  |\varphi_0\rangle|0_\mathrm{P}^\prime\rangle\hat{\varrho}
  \langle0_\mathrm{P}^\prime|\langle\varphi_0|
  \hat{X}_4|\varphi_0\rangle\right)
  \nonumber\\&=&
  \frac{1}{p(1_{b_0})}\sum_{m,n=0}^N
  \frac{\langle m|\hat{R}_a^\dagger(\Phi)\hat{R}_a(\Phi)|n\rangle}{N+1}
  \;|m\rangle\langle m|\hat{\varrho}|n\rangle\langle n|,
\label{rho0fig4}
\end{eqnarray}
where $p(1_{b_0})$ $\!=$
$\!(N+1)^{-1}\sum_{k=0}^N\langle k|\hat{R}_a^\dagger(\Phi)
\hat{R}_a(\Phi)|k\rangle\langle k|\hat{\varrho}|k\rangle$
and furthermore $\hat{R}_a(\Phi)$ $\!:=$
$\!\mathrm{e}^{\mathrm{i}\Phi\hat{a}^\dagger\hat{a}}$ $\!\hat{R}_a$
$\!\mathrm{e}^{-\mathrm{i}\Phi\hat{a}^\dagger\hat{a}}$.
We see that $\hat{\varrho}_{\mathrm{red}}$ still depends on the input state.
In particular, for $T_k=1$, (\ref{rho0fig4}) turns into (\ref{rho0fig3}).
On the other hand, the input can even pass the right converter without
disturbance, $\hat{\varrho}_{\mathrm{red}}$ $\!=$ $\!\hat{\varrho}$, if
$\hat{R}_a(\Phi)$ is a projector onto the state $|0_\mathrm{P}\rangle$,
$\langle m|\hat{R}_a(\Phi)|n\rangle$ $\!=$
$\!\langle\varphi_m|\hat{U}_\Phi\hat{R}_b\hat{U}_\Phi^\dagger|\varphi_n\rangle$
$\!=$ $\!(N+1)^{-1}$.
However, in contrast to figure~\ref{Fig3}, the reduced state at the single-mode
output of the left-hand converter,
\begin{eqnarray}
  \hat{\varrho}_{\mathrm{red}}^\prime
  &:=&\mathrm{Tr}_{\mathcal{H}_b}
  \left(\hat{M}^\prime|
  \varphi_0\rangle|0_\mathrm{P}^\prime\rangle
  \langle0_\mathrm{P}^\prime|\langle\varphi_0|
  \hat{M}^{\prime\,\dagger}\right)
  \nonumber\\&=&
  (N+1)^{-1}\hat{P}_{ba^\prime}^\dagger\hat{P}_{ba^\prime},
\label{rho1fig4}
\end{eqnarray}
is now independent of the input state $\hat{\varrho}$.
Bob is left with ``white noise'' if Alice refuses to communicate with him.
%-----------------------------------------------------------------------------
\subsection{Unconditional operation}
\label{sec6.2}
%-----------------------------------------------------------------------------
Until now, we have assumed that Alice sends a classical trigger bit to Bob in
case she has detected photon presence in channel $b_0$ as well as a given value
of the phase $\Phi$ defined by the construction of the entangled state source
and Bob then opens the shutter S.
Consider now an alternative situation in which, after detection of photon
presence in some channel $b_k$, Alice uses the classical channel to tell Bob
the respective channel number $k$ and phase value $\tilde{\Phi}$ measured.
(\ref{ytela'a}) is then generalized to
\begin{eqnarray}
  \hat{Y}_{\mathrm{tel}}(1_{b_k},\tilde{\Phi})
  &=&\langle\varphi_k|\langle\tilde{\Phi}_\mathrm{P}|
  \hat{M}^\dagger\hat{U}_\Phi\hat{A}^\dagger\hat{M}^\prime
  |0_\mathrm{P}^\prime\rangle|\varphi_0\rangle
  \nonumber\\&=&
  (N+1)^{-1}\hat{P}_{ba^\prime}^\dagger\hat{A}_b^\ast
  \hat{U}_{k\tilde{\Phi}}
  \hat{P}_{ba},
\label{ytela'ak}
\end{eqnarray}
where $\hat{U}_{k\tilde{\Phi}}$ $\!=$
$\!\hat{U}_\Phi\hat{V}_b^k\hat{U}_{\tilde{\Phi}}^\dagger$ depends on the values
$k$ and $\tilde{\Phi}$ detected. On Bob's side, the shutter S is replaced with
state converters described in section (\ref{sec3.2}). Bob uses these to convert
his respective output state $\hat{\varrho}^\prime$ into its multi-mode
counterpart, to which a unitary transformation
$\hat{U}_{k\tilde{\Phi}}^\dagger$ is applied by means of a beam splitter array,
and eventually the transformed state is converted back into its single-mode
counterpart. In order to describe the transformation of Alice's state
$\hat{\varrho}$ into Bob's modified state $\hat{\varrho}^{\prime\prime}$
according to (\ref{yomega}), we use in place of (\ref{ytela'ak}) an effective
transformation operator
\begin{eqnarray}
  \hat{\Upsilon}_{\mathrm{tel}}(1_{b_k},\tilde{\Phi})
  &=&(N+1)\hat{P}_{ba^{\prime\prime}}^\dagger
  \hat{U}_{k\tilde{\Phi}}^\dagger
  \hat{P}_{ba^\prime}
  \hat{Y}_{\mathrm{tel}}(1_{b_k},\tilde{\Phi})
  \nonumber\\&=&
  \hat{P}_{ba^{\prime\prime}}^\dagger
  \hat{U}_{k\tilde{\Phi}}^\dagger
  \hat{A}_b^\ast
  \hat{U}_{k\tilde{\Phi}}
  \hat{P}_{ba},
\label{yteleffa'a}
\end{eqnarray}
which depends on $k$ and $\tilde{\Phi}$ if
$[\hat{A}_b^\ast,\hat{U}_{k\tilde{\Phi}}]$ $\!\neq$ $\!\hat{0}$.
In particular, if the beam splitters in dashed parentheses in figure~\ref{Fig4}
are removed, $T_k=1$, then at each trial, one of the ON/OFF-detectors D clicks,
$p(1_b)=1$. A well known example is the unconditional teleportation, which is
achieved by removing also the beam splitter array $\hat{U}$, so that
(\ref{yteleffa'a}) reduces to
$\hat{\Upsilon}_{\mathrm{tel}}(1_{b_k},\tilde{\Phi})$ $\!=$
$\!\hat{P}_{ba^{\prime\prime}}^\dagger\hat{P}_{ba}$.
%-----------------------------------------------------------------------------
\newpage\section{Conclusion}
\label{sec7}
%-----------------------------------------------------------------------------
We have suggested a cross-Kerr interaction based device allowing conversion
between single-mode states truncated at some photon number $N$ and ($N$+1)-mode
states whose total photon number is one. 
As possible applications with regard to single-mode states, we have considered
the implementation of unitary and non-unitary transformations, overlap
measurements with orthogonal and non-orthogonal sets of states, and
telemanipulation.

Throughout the work we have distinguished between a conditional and an
unconditional mode of operation. Whereas the unconditional mode is based on
detection and preparation of Pegg--Barnett phase states, the conditional mode
may apply arbitrary states with non-zero Fock expansion coefficients.
For example, if the outgoing $a$-mode of device $\hat{M}$ in
figure~\ref{Fig2}(a) passes a separate (highly transmittive) beam splitter,
whose second input port is prepared in a (strong) coherent state, then
detection of photon absence in the first output port of this beam splitter by
means of an ON/OFF-detector approximates the detection of a truncated coherent
state, since the incoming $a$-mode of device $\hat{M}$ is by definition
prepared in a photon number truncated state. Within our work, attention has
however been limited to Pegg--Barnett states (as well as lossless devices and
perfect mode-matching), thus allowing a unified depiction of the principle.
%-----------------------------------------------------------------------------
\section*{Acknowledgement}
%-----------------------------------------------------------------------------
This work has been supported by the Deutsche Forschungsgemeinschaft.
%-----------------------------------------------------------------------------
\begin{appendix}
%-----------------------------------------------------------------------------
\section{Proof of (\ref{tkrkconseq})}
\label{app1}
%-----------------------------------------------------------------------------
Let $|F\rangle_{b_k}$ $\!=$ $\!\hat{F}(\hat{b}_k^\dagger)|0\rangle_{b_k}$
be an arbitrary single-mode state. From (\ref{tkrk}) it follows that
\begin{eqnarray}
  _{c_k}\langle0|\hat{U}_k^\dagger|0\rangle_{c_k}|F\rangle_{b_k}
  &=&_{c_k}\langle0|\hat{U}_k^\dagger\hat{F}(\hat{b}_k^\dagger)\hat{U}_k
  \hat{U}_k^\dagger|0\rangle_{c_k}|0\rangle_{b_k}
  \nonumber\\&=&
  _{c_k}\langle0|\hat{F}(T_k^\ast\hat{b}_k^\dagger+R_k^\ast\hat{c}_k^\dagger)
  |0\rangle_{c_k}|0\rangle_{b_k}
  \nonumber\\&=&
  \hat{F}(T_k^\ast\hat{b}_k^\dagger)|0\rangle_{b_k}
  \nonumber\\&=&
  T_k^{\ast\,\hat{b}_k^\dagger\hat{b}_k}\hat{F}(\hat{b}_k^\dagger)
  |0\rangle_{b_k}
  \nonumber\\&=&
  T_k^{\ast\,\hat{b}_k^\dagger\hat{b}_k}|F\rangle_{b_k},
\end{eqnarray}
and therefore
\begin{equation}
  _{c_k}\langle0|\hat{U}_k^\dagger|0\rangle_{c_k}
  =T_k^{\ast\,\hat{b}_k^\dagger\hat{b}_k}
\end{equation}
holds in general. The adjoint of this equation just yields (\ref{tkrkconseq}).
%-----------------------------------------------------------------------------
\end{appendix}
%-----------------------------------------------------------------------------

%-----------------------------------------------------------------------------
\end{document}